北京大学物理百年华诞纪念专刊 · 评述

# 压缩重子物质: 从原子核到脉冲星

徐仁新*

北京大学物理学院天文学系, 核物理与核技术国家重点实验室, 北京 100871
*联系人, E-mail: r.x.xu@pku.edu.cn

**摘要**　　尽管物质世界以不熟悉的暗物质、暗能量为主, 但自然界的精彩却要归功于相对少量的那些重子物质. 日常生活中原子核之间 Coulomb 排斥有效地阻止它们通过挤压物质而聚合起来, 然而天体极端环境时常出人意料: 大质量恒星演化至晚期时, 其核心引力如此之强以至于其他任何力(当然包括 Coulomb 排斥)都难以媲美——压缩重子物质就这样在超新星爆发过程中诞生, 并表现为观测到的脉冲星. 对这类压缩重子物质的研究将不仅加深对强作用基本性质的认识, 而且有助于检验引力理论、探测低频引力波、建立精确的时间标准和导航体系, 还是我国在建或拟建大型天文望远镜的核心课题. 历史上 Landau 曾推测恒星核心存在巨大的原子核(后来发展成为"中子星"模型), 我们经过研究后却认为脉冲星其实是由夸克集团而构成的凝聚体. 夸克集团作为超新星爆发形成压缩重子物质的基本单元这一想法十年来并未被观测排除, 而且我们还期待未来功能更强大的观测设备检验它.

**关键词**　　脉冲星, 核物质, 夸克物质, 原子核

**doi:** 10.1360/132013-347

　　诚然物质世界 95%以上由不熟悉的暗物质、暗能量组成, 但是我们这个丰富多彩的世界却要归功于剩余的重子物质; 后者主要由夸克(并伴随轻子)构成, 已被人们透彻地研究. 夸克间强作用的渐近自由性质使得其在宇宙极早期处于游离状态(夸克胶子等离子体, QGP); 随着宇宙膨胀、降温而导致夸克之间耦合增强, 如下两个因素有助于色单态重子在年龄几十微秒的宇宙中诞生(零重子数的介子最终衰亡): 强作用存在三种色荷, 重子之间可能存在排斥芯(经验地讲, 至少核子如此). 质量最轻的重子(即核子)得以残存, 直到宇宙年龄百余秒时部分核子又重新聚合(主要产物为 $^4$He). 自然界中绝大多数比较重的原子核是在跟恒星相关的过程中合成的. 目前发现重子数最大的核是 118 号元素(Ununoctium); 它是通过加速器中重元素核反应合成的, 重子数为 294.

　　不过, 本文将逆向于以上的思路: 可否大规模地将当今宇宙中的原子核再次聚合起来? 鉴于荷电轻子(电子质量最轻, 也最稳定)不参与强作用, 电子在利用较弱的电磁作用与原子核这个强作用客体耦合在一起构成原子时, 其分布范围远远超过核. 原子中

外层电子一般非相对论性运动, 因而原子典型大小要远大于电子 Compton 波长(局域相对论电子的特征尺度, $\lambda_c$=0.024 Å); 而原子核的大小却为飞米量级, 远小于原子大小. 这样, 原子核之间强烈的 Coulomb 排斥就能有效地阻止人们通过挤压物质而让它们聚合起来. 然而天体的极端环境常常出人意料: 大质量恒星演化至晚期时, 其核心引力如此之强以至于其他任何力(当然包括 Coulomb 排斥)都难以媲美. 压缩重子物质就这样在超新星爆发过程中再次诞生!

这样的压缩重子物质将是本文关注的焦点. 它有什么性质? 为何值得人们努力研究? 较于极早期宇宙中出现的 QGP, 超新星残留压缩重子物质的温度要低得多; 前者高于 150 MeV, 而后者低于 50 MeV(往往只有~keV). 如此看来, 冷的压缩重子物质倒有点类似原子核, 只不过重子数(~$10^{57}$)极高罢了. 那么, 它会跟熟悉的原子核有何区别? 具体而言, 压缩重子物质内还是两味价夸克吗? 夸克是类似于核子内那样聚集的还是自由的? 等等. 本文拟试图回答这些问题. 我们将尽力向读者阐明如下观点: 压缩重子物质内的价夸克是三味的, 它们是成团而存在的; 每个夸克集团内的夸克数目可能是 6, 12, 18 等; 压缩重子物质的物态很硬并会处于有刚性的固态, 其观测上表现为脉冲星; 除了认识低能强作用特征方面的意义, 脉冲星研究还有助于检验引力理论、探测低频引力波、建立精确的时间标准和导航体系等, 而且是我国在建或拟建大型望远镜的关键课题.

## 1 历史的回顾: 从原子到色超导

如前所述, 超新星爆发后形成的脉冲星由压缩重子物质构成, 可以形象地看做"大原子核". 在具体考察这类大原子核之前, 先让我们回顾构成普通物质的小原子核. 古代哲学家提出原子概念后人们一般认为原子是构成物质的基本粒子, 直到 1897 年 Thomson 在研究阴极射线时发现了原子中存在极轻的带负电粒子——电子. Thomson 用"葡萄干布丁"图像形象地诠释他的原子模型: 一些带负电的电子镶嵌于均匀的正电荷背景上. 1911 年, 这一模型受到了 Rutherford 等人利用 $\alpha$ 粒子在金箔上散射实验的冲击: 原子(大小~$10^{-8}$ cm)的质量及其正电荷实际上集中于一极小的区域(~$10^{-13}$ cm)! 鉴于原子核如此之小, Fowler[1]于 1926 年推测: 仅受原子核和电子自身体积所限, 密度达到 $10^{14}$ g/cm$^3$ 的物质并非不可能存在. 这是关于压缩重子物质最早有文献记载的、带有猜测性的思考. 尽管 1920 年 Rutherford 就推测"由电子和质子紧密结合的"中性双子(Neutral Doublet, 后来称之为中子)和质子是原子核的基本组份, 但直到 1932 年发现中子之后关于原子核组份的才终有定论.

那么, 小原子核感觉起来像个什么样子呢? Gamow 于 1928 年至 1931 年在哥本哈根大学访问的时候曾天才般地提议将原子核视为由"核物质"构成的不可压缩液滴[2]. 这一图像影响深远, 即便有人在 1974 年基于原子核巨共振现象与弹性体的相似性而认为原子核呈固态[3], 液滴模型依然流行.

为了理解恒星的能源, 1932 年(注: 在中子发现之前[4])Landau 推测在恒星的中心存在"大原子核"[5]: "We expect that this must occur when the density of matter becomes so great that atomic nuclei come in close contact, forming one gigantic nucleus." 他认为大原子核的存在导致恒星发光: 起先怀疑量子力学、能量守恒等基本物理定律在大原子核区失效[5], 后来探讨形成大原子核过程中所释放出的引力能就是恒星能量的根本来源[6]. 尽管现在知道这一看法是错误的(后来确认恒星发光源于核聚变能[7]), 但确实是 Landau 首次比较现实地意识到可能存在密度与原子核相当的宏观物质(这可看作 Gamow 所言微观"液滴"概念的推广). 值得一提的是: 尽管以凝聚态物理特别是超流理论方面的基础工作而著名, Landau 也很看重"大原子核"这一想法. Landau 一生在 *Nature* 杂志共发表六篇文章[8], 但三篇单独署名文章之一的文献[6]就是基于文献[5], 而另外两篇与他获得 Nobel 奖有关的文章只是先前工作的总结. 文献[6]也被 Kapitsa 用以营救在监狱中的 Landau[4].

如果把质子(p)与中子(n)都视作基本粒子的话, 那么 Landau 提出的宏观大原子核一定是丰中子的, 是"中子星"概念的原型. 这是因为电子(e)将处于大原子核的内部(而电子在小原子核内出现的几率甚微), 参与如下化学平衡

$$e+p \leftrightarrow n+\nu_e, \qquad (1)$$

在这里中微子($\nu_e$)的化学势可以忽略. 对于质量远低于核子的电子而言, 其化学势可以通过形成中子而降低. 可见, 如果核子像电子和中微子那样是基本的、无结构的"点"粒子, 电子在大原子核内参与弱平衡必然出现富中子. 这就很容易理解, 大原子核的概

念在中子发现后很快被"中子星"一词所取代. 然而问题是: 包括质子、中子在内的强子并非如 1930 年代想象的那样基本, 它们其实是有结构的!

除了思辨之外, 是否存在中子星的天文观测迹象呢? 1934 年, Baade 与 Zwicky[9]研究源于超新星的宇宙线(离子), 并提出形成中子星过程所释放出的能量可能是超新星而非一般恒星的能量来源. 然而他们对于中子星形成机制的看法("If neutrons are produced on the surface of an ordinary star they will 'rain' down towards the center if we assume that the light pressure on neutrons is practically zero")并非正确. 后续还有人讨论在超新星核心区域会形成热中子星而向外辐射 X 射线[10], 并且认为中子星所拥有的转动能会为超新星遗迹供能[11]. 1967 年, 射电脉冲星的发现是一个突破[12], 且很快被证认为旋转中子星[13].

尽管考虑到超子自由度后人们迅速探讨了核心区域的超子成分, 但关于中子星组成方面的较大改进却源于 1960 年代强子结构模型的进展. 1964 年 Gell-Mann[14]和 Zweig[15]分别提出强子由更基本的夸克所组成. 1969 年 Ivanenko 和 Kurdgelaidze[16]探讨极致密的中子星核心可存在自由夸克. 1970 年 Itoh[17]尝试计算了 u, d, s 三味自由夸克构成星体的流体力学平衡. 1971 年 Bodmer[18]探讨了带奇异数塌缩原子核(Collapsed Nuclei)的稳定性. 真的存在远多于 3 个自由夸克组成的客体吗? 以上这些带猜测性的研究在 1973 年 Gross 和 Wilczek[19], Politzer[20]证明了夸克之间色作用的渐近自由特性后变得现实. 1984 年 Witten[21]分析了奇异夸克物质的稳定性并探讨了可能的天体物理后果: 宇宙早期相变、宇宙线、奇异星. 1986 年 Haensel 等人[22]和 Alcock 等人[23]计算给出了奇异星的结构并讨论了可能的观测表现.

现在我们知道, 夸克之间相互作用由强作用主导, 描述该作用的基本理论是量子色动力学(QCD). 基于渐近自由, 低温高密下无疑可能会出现夸克 Fermi 气; 考虑到夸克之间较弱的色作用, 1998 年 Alford, Rajagopal 和 Wilczek[24]提出 Bardeen-Cooper-Schrieffer(BCS)类色超导夸克物质. 尽管微扰 QCD 成功地描述了若干高能现象, 但低能 QCD 却是人们遇到的重大挑战. 如果几倍核物质密度时夸克之间的色相互作用还比较强(强耦合参数 $\alpha_s \geqslant 1$), 夸克 Fermi 气态可能是不稳定的, 但这并不意味着夸克一定因禁在强子内以便夸克自由度可以忽略; 鉴于奇异夸克在致密物质中较容易激发且 QCD 原则上允许构成多夸克态, 我们提出在致密星内部可能会形成 u, d, s 三味的夸克集团[25]. 诚然, 类似于核子, 夸克集团间还可能存在着剩余的短程排斥(以及远程吸引)相互作用; 这使得它们可看作经典粒子: 当温度足够低于集团间作用能时, 最终形成固体夸克集团物质.

总之, 目前有两种思潮左右着人们认识脉冲星的内部结构. 一种认为脉冲星以像中子、质子这些仍然未被"压碎"的强子为主组成(即通常的中子星模型). 强子星、混合/混杂星就属于这一类. 另一种认为强子内的夸克都被"压出来"而形成自由态(即夸克星模型). 而我们提出的"夸克集团星"则介于该两极端观念之间(图 1), 类似强子的夸克集团是其基本成分.

## 2 夸克集团态: 微观描述

超新星爆发而产生的压缩重子物质本质如何? 这是涉及低能强相互作用(非微扰 QCD)的一个基本问题, 其根本解决关系到克雷数学研究所提出的千禧七大疑难之一(即"Yang-Mills Theory"). 从第 1 节的介绍我们知道, 在人类认识亚核子之前 Landau 首次推测这类物质的存在. 后来天文学的进展使得这一原始的看法逐步改进, 发展成为目前流行的"中子星"概念. 诚然, 人类对于微观的认识已经进入"夸克和轻子"时代; 从现代物理角度来看, Landau 当年的出发点是否还靠得住呢? 让我们就从分析 Landau 当年所犯的两个错误开始.

### 2.1 Landau 在 23 岁时所犯的两个错误

受 20 世纪 30 年代科学发展背景所限, Landau 于 1932 年所撰一文[5]中关于大原子核的看法存在两个错误的观念. 下面我们就来讨论和分析之.

错误 1: 大原子核内量子力学失效以致质子和电子可"非常紧密地结合"(用现在的话讲就是束缚于飞米量级的尺度上), 即形成 Rutherford 于 1920 年所猜测的"双子". 当然我们现在知道, 电子与质子间由电磁相互作用束缚; 若服从被已经很好地检验的量子力学的话, 那么它便只能构成氢原子那样的相对弱束缚系统(在埃的量级上).

Landau 之所以会犯这样的错误是因为那时尚未发现弱相互作用. 我们现在知道, 弱相互作用能够通

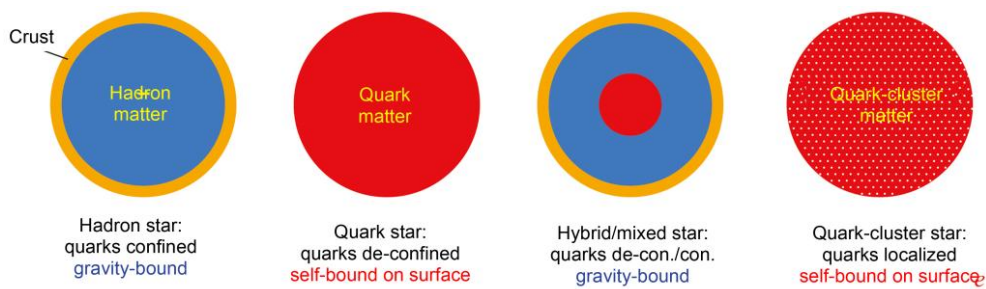

**图 1**  (网络版彩图)不同类型的脉冲星结构模型. 强子星(Hadron Star)内部无游离夸克组成的夸克物质, 而中心密度足够高时会形成具有夸克物质核心的混合/混杂星(Hybrid/Mixed Star). 几乎完全由夸克物质构成的致密星称为夸克星(Quark Star). 以夸克集团为主要自由度的夸克集团星(Quark-Cluster Star)有别于通常的中子星和夸克星. 强子星和混合/混杂星是引力束缚的, 这两类致密星都拥有由原子核和电子气等组成的壳层(Crust); 而夸克星和夸克集团星的表面是色作用自束缚的, 原则上可以不具有壳层. 表面直接裸露的夸克(集团)星称为裸夸克(集团)星

**Figure 1**  (Color online) Different models of pulsar structure. Quarks are confined in hadrons of hadron stars, while quark matter would exist in a core of hybrid/mixed star whose central density could be high enough to make quarks de-confined. A quark-cluster star is a condensed object of quark clusters via residual color interaction, which distinguishes from conventional both neutron and quark stars. Both hadron star and hybrid/mixed star are bound by gravity, which are covered by crusts with nuclei and electrons, while quark star and quark-cluster star are strongly self-bound on surface. A bare (crusted) quark or quark-cluster star is one without (with) a crust.

过反应方程(1)将一个质子和一个电子转变为一个新的粒子, 即稍后由 Chadwick 发现的"中子".

在 20 世纪已成功建立的粒子物理标准模型中, 基本费米子(夸克与轻子)及传播其间相互作用的规范玻色子才是没有结构的, 而质子(uud)与中子(udd)并非 30 年代人们所认为的点粒子. 所以方程(1)的本质其实是通过反应 $e+u \leftrightarrow d+\nu_e$ 将一个 u-夸克转变为一个 d-夸克. 实际上, 除了核子中含有的 u 和 d 价夸克外, 总共有 6 味夸克, 并可以根据质量差异分为两组: 质量低于 GeV 的轻味夸克{u, d, s}以及质量超过 GeV 的重味夸克{c, t, b}. 尽管味守恒会在强相互作用下成立, 但弱相互作用无疑会改变夸克的味.

这就进一步产生一个发人深省的问题: 既然包括奇异夸克在内的轻味夸克比重味夸克容易激发产生, 那么大原子核是否奇异数非零呢? 其实在几倍核物质密度情形下, 无论是据 Heisenberg 关系估计的零点能还是简单 Fermi 气图像所给出的 Fermi 能都大于轻夸克之间的质量差($m_u \sim m_d \sim$ 10 MeV, 而 $m_s \sim$ 100 MeV). 然而, 为何原子核系统只具有{u,d}两味而不是{u,d,s}三味对称性呢? 这可能与微观原子核和宏观"大"原子核的尺度有关. 微观原子核半径远小于电子 Compton 波长, 电子处于原子核外, 动能 $E_e^{\text{mac}}$ 与势能相当, 都是可忽略的,

$$E_e^{\text{mic}} \sim e^2/l \sim \alpha_{\text{em}}^2 m_e c^2 \sim 10^{-5} \text{MeV}, \quad (2)$$

这里 $l$ 是电子分布的空间尺度, $\alpha_{\text{em}} \simeq 1/137$ 为电磁作用耦合常数. 而对于尺度远大于电子 Compton 波长的大原子核而言, 电子处于其内部, 动能 $E_e^{\text{mic}}$ 不可忽略

$$E_e^{\text{mac}} \sim \hbar c n^{1/3} \sim 10^2 \text{MeV}, \quad (3)$$

这里 $n$ 为电子的数密度. 若大原子核具有三味对称性(而非一般原子核研究中所议"同位旋对称性"), 则可显著降低电子对系统总能量的贡献. 因此, 宏观的压缩重子物质很可能是三味对称的; 这可以看作对 Witten 猜想[21]的推广.

错误 2: 引力是决定大原子核一般属性和整体结构的唯一相互作用, 因而 Landau 认为引力能释放主导, 且大原子核是引力束缚的. 他曾于 1937 年又写道: "Thus we can regard a star as a body which has a neutronic core the steady growth of which liberates the energy which maintains the star at its high temperature; the condition at the boundary between the two phases is as usual the equality of chemical potential" (译文: 我们因而认为恒星应该具有一个中子核心, 其稳恒的增长释放着维持恒星高温的能量; 这两个相的边界条件还是由通常的化学势平衡确定).

Landau 犯下这第二个错误是因为当时未发现强相互作用. 强相互作用至少会为大原子核研究带来两方面的重要影响: (1) 核能可以通过核反应释放出来, 因此后来意识到核能才是恒星能量的真正来源;

(2) 如同一般原子核那样, 大原子核也可能是通过强相互作用而自束缚的, 而非仅仅引力束缚; 自束缚表面甚至可以直接裸露而不具有由原子核和电子组成的壳层和大气层.

值得一提的是: 核力作为有别于电磁力、引力之外的一种特殊力早在 Gamow 关于原子核的液滴图像中就已经被意识到了. 因缺乏核合成释能的明确实验根据, 很遗憾 Landau 先验地认为引力能的释放才会是显著的. 随着核反应实验的进展, Bethe 提出了氢核聚变成氦核的 CNO 循环机制[7]. 由此看来, 关于恒星能源问题, Landau(1932)和 Bethe(1939)一个错误、一个正确的解也跟核力的认识有关: Landau 忽略了核力的重要性因而祈求于"大原子核"贡献的引力能, 而 Bethe 在核反应放能的启示下找到恒星能源问题的正确答案.

认识到存在强相互作用的另一个后果是: 超新星爆发而形成的压缩重子物质可能因强作用而自束缚. 白矮星是引力束缚主导的, 这体现在如下两个方面: (1) 星体中心密度远高于表面密度; (2) 质量愈大的星体引力愈强, 其半径也就愈小. 主序恒星也是引力束缚的, 只是物态有别于白矮星. 如果脉冲星也是引力束缚的, 则其质量-半径关系应该类似于白矮星, 只不过极限质量有所增加且半径要小得多罢了. 然而, 如果脉冲星是强作用自束缚的, 类似于原子核中的核子因核力而束缚, 则质量-半径将显著不同于引力束缚的白矮星: 星体的质量越大往往其半径也越大; 并且可以允许存在近乎无引力(Gravity-free)的小质量脉冲星类天体(小原子核可以看作特殊的无引力系统, 具有统一的密度——饱和核物质密度).

总之, 对于强相互作用与弱相互作用知识的缺乏导致 Landau 在 20 世纪 30 年初代犯下了两个错误. 80 多年后的今天, 强力与弱力在核物理、粒子物理以及天体物理中都起到了相当重要的作用, 大原子核概念的发展和修正无疑势在必行. 我们将做如下面尝试.

## 2.2 改进的大原子核概念

诚然大原子核的概念逐步发展成为现代中子星模型, 但大、小原子核之间确实存在基本差别. 如前所述, 它们有两个关键不同点. (1) 由于大原子核的尺度远大于小原子核, 电子将存在于大原子核的内部; (2) 由于引力在大原子核中的作用远强于在小原子核中, 大原子核的平均密度一般要略要高于小原子核的密度, 达几倍核物质密度. 相较于小原子核而言, 这两点不同会使大原子核表现出一些特别之处. 关于大原子核, 人们很容易会提出这样的问题: 那里的夸克依然如重子情形是三个成一集团吗? 大原子核会否有奇异数? 还处于液态吗? 经过多年研究我们认为: 为了更好地理解脉冲星类致密天体的各种观测表现, 大原子核其实是由以带奇异数夸克集团为基本单元凝聚而成的固态物质组成的[25]. 我们将在第 3 节进一步阐述.

致密星内部的真正物态跟 QCD 相图(以温度 $T$ 和重子化学势 $\mu_B$ 为变量)密切相关, 后者在核物理与粒子物理界一直是个热门话题. 在低温情况下, 由于渐近自由性, 重子密度过高的致密物质会从强子相过渡到夸克解禁相. 但一个极其严肃的问题是: 在实际的致密星内部, 重子数密度是否会足够高(低)以至于能使夸克解禁(囚禁)呢?

让我们从如下两个方向来考察这一基本问题.

一个方向是从强子态着手(自下而上的考虑). 我们先假设在脉冲星中的夸克是非解禁的. 但是由于刚才谈到的两点不同, 这样的非解禁态并不见得能天真地视为通常的强子物质态. 随着大原子核的尺度与密度的增加, 轻夸克味对称性可能会逐渐得到恢复而终形成 u, d 与 s 三味夸克数密度近乎相等的奇异物质(但夸克并非一定处于游离态). 我们知道核子是最轻的重子; 但如果大原子核内质子和中子数还像小原子核内那样近似相等的话, 电子的典型能量将会达到 $10^2$ MeV 的量级(这一能量已经接近甚至大于 s-夸克与 u/d-夸克间的质量差). 然而, 电子数目和能量在奇异物质中却是可忽略的. 即使考虑到强相互作用的存在, 具有轻味对称性的大原子核也可能是冷致密物质的基态. 进一步地, 80 年多前 Landau 就已注意到, 因引力能的释放大原子核会倾向稳定. 因此, 在大原子核中或许三味轻夸克能形成一种类似于重子的集团态. 我们将这种新的强相互作用态称为"夸克集团". 由于它们的质量比较大以及之间的强作用, 夸克集团在低温下会呈非相对论运动并且可能被局限于晶格附近, 最终形成以夸克集团为基本单元的固体.

另一个方向是从自由夸克态着手(自上而下的考虑). 在有关低温超核密度物质的物相研究中, 类似于 BCS 色超导相被关注. 如果在现实的致密星中夸

克之间的耦合确实较弱, 以致可用微扰 QCD 理论进行处理的话, 那么色超导研究便是值得的. 然而, 微扰 QCD 的结论至少在能标 1 GeV 以上才是可信的; 但在实际脉冲星(典型质量约为 $1.4M_\odot$, 半径约为 10 km)的内部, 夸克的化学势却只有大约 0.4 GeV. 所以致密星中的强相互作用很可能会很强以使夸克成团. 记夸克集团尺度为 $l_q$, 夸克之间存在类似 Coulomb 势的色相互作用且作用能为 $E_q$. 对于组份夸克质量 $m_q \simeq 300$ MeV 的夸克集团, 可以用 Heisenberg 关系估算这两个量:

$$l_q \sim \frac{1}{\alpha_s} \frac{\hbar c}{m_q c^2} \simeq \frac{1}{\alpha_s} \text{fm}, \quad E_q \sim \alpha_s^2 m_q c^2 \simeq 300\alpha_s^2 \text{ MeV}. \quad (4)$$

如果相互作用常数 $\alpha_s>1$, 那么 $E_q$ 就会接近甚至超过大小约为 0.4 GeV 的化学势; 这对于夸克的费米气图像而言是极其不利的. 而事实上, Dyson-Schwinger 方程方法得到几倍核物质密度下夸克之间的色耦合还非常强[26], 可达到 $\alpha_s \gtrsim 2$. 因此, 致密星内部的夸克很可能成团并局域化.

是故, 我们猜想在 QCD 相图上会存在夸克集团相(图 2). 夸克集团相在高温时为液态, 但在热运动动能低于夸克集团间的剩余强相互作用能量时变为固态. 实际的脉冲星温度较低; 因此, 作为对 Landau 大原子核图像的修正, 可以认为脉冲星即为固态夸克集团星. 夸克集团星与大铁球较为相似, 只是后者晶格上是离子/原子核、电磁相互作用主导, 前者则为夸克集团、强相互作用为主. 值得注意的是, 由于轻味对称性的破缺, s-夸克的数密度要稍少于 u/d-夸克. 这样奇异夸克集团物质中便存在电子和强电场, 使得夸克集团星的表面原则上允许存在由正常物质构成的壳层.

夸克集团会是什么样的呢? 我们知道Λ粒子(组成为 uds)是满足轻夸克味对称性的, 人们或许可以认为夸克集团就类似于Λ粒子. 然而, Λ粒子之间的二体相互作用却为吸引势, 这会使更多的夸克聚集在一起. 受到最近 H 双重子(组成为 uuddss)格点 QCD 模拟的启发, 我们考虑一种由 H 集团构成的夸克集团星[27]. 除了能够解释一般的致密星表现外, H 集团星(或者更简单地称之为 H 星)模型在一定的参数下还能给出极限质量高于 $2M_\odot$ 的结果.

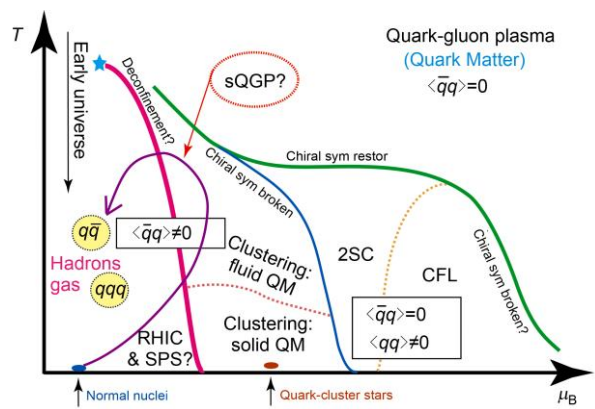

**图 2** (网络版彩图)推测的 QCD 相图. 纵坐标 $T$ 为温度, 横坐标 $\mu_B$ 为重子化学势或重子数密度. 在低温和低密时, 夸克囚禁在强子内; 而足够高温或高密时因渐近自由而产生游离夸克, 为夸克物质或夸克胶子等离子体态. 在几倍核物质密度下, 即现实的致密星内部, 夸克之间的耦合可能还很强, 从而形成夸克集团. 低温夸克集团物质将处于固态

**Figure 2**   (Color online) A QCD phase diagram contains the conjectured quark-cluster phase. $T$: temperature, $\mu_B$: baryon chemical potential which can also be considered simply as the number density of quarks/baryons. Quarks are confined in hadrons when both density and temperature are low, while they are free when both density and temperature are very high. Nevertheless, a quark-cluster state is conjectured for cold dense matter at a few nuclear densities where chiral symmetry could still be broken so that the mass of dressed quark would be order of 300 MeV. This kind of quark-cluster matter should be in a solid state at low temperature, as a rigid body.

## 3  夸克集团星: 天体表现

图 1 展示了不同的脉冲星结构模型; 这些模型所预言的不同观测特征及其观测检验无疑将受到天体物理学者的关注. 下面我们就以一般中子星(其原型为旧版的大原子核, 但有所发展)和固态夸克集团星(改进版的大原子核)为示例来分析. 从脉冲星类天体的观测上来看, 它们之间最大的区别在哪里呢? 我们认为在于如下两点. (1) 表面: 在中子星表面, 包括原子核和电子在内的粒子被引力束缚住; 而对于夸克集团星, 表面的夸克和电子都被有效地自束缚着(夸克被强力而电子被强电磁力束缚). (2) 整体结构: 一方面描述星体结构的物态方程的软硬会预言两类致密星不同的极限质量, 另一方面固态夸克集团星具有整体刚性而中子星却没有. 这些差异会体现在观测表现上, 有助于各类模型的观测检验[26–29]. 过去一些相关研究归纳如下.

表面束缚方式的差异体现为如下两点: 星体的质量-半径(M-R)关系与表面粒子束缚能. 自束缚的夸克集团星的表面密度(>$10^{14}$ g/cm$^3$)是非零的, 未达到极限质量时其半径会随着质量的增加而增加(特别地, 对于低质量夸克集团星, 其引力作用可以忽略, 可以认为 $M \propto R^3$); 而对于引力束缚的中子星而言, 其半径会随着质量的增加而减小(质量越大引力越强, 则对星体的"压缩"就越厉害, 故而半径越小). SAXJ1808.4-3658 独特的质量-半径关系可能表明它是一颗夸克星[30]. 不过, 更直观地反映表面性质的不同应该体现在表面粒子束缚能的差异上(下段阐述).

射电脉冲星是脉冲星类天体的主体. 射电子脉冲普遍存在飘移现象, 该观测特征可以较自然地在 Ruderman-Sutherland 模型(RS 模型)框架内理解. 然而, RS 模型依赖于极冠区较高的粒子束缚能以产生间隙火花放电. 1970 年代初对强磁场中物质束缚能的估计表明中子星表面离子的束缚能大于 10 keV, 因而可以在其极冠区形成火花放电过程; 但 1980 年代及后续更详细的计算显示中子星表面离子的束缚能往往不能足够高到产生火花放电, 除非表面磁场接近甚至超过临界磁场($B_q$=4.14×$10^{13}$ G). 但在裸夸克集团星框架内, 由于其表面粒子强的自束缚, 这一问题得到自然解决. 除此之外, 强的表面束缚还能允许存在极大光度的爆发现象: 夸克集团星的辐射光度不会受到 Eddington 极限的限制, 因而以夸克集团星为中心致密天体的超新星与γ射线暴均可能是光子驱动的. 再者, 未探测到无磁层活动的"死亡"脉冲星热辐射谱中的原子谱线或表明由中子星模型所预言的大气层并不实际存在, 这一观测特性也与裸夸克集团星表面属性相符.

让我们再来看看这两类致密星整体结构上的不同. 第 2 节中已经提到, 夸克集团星整体处于固态, 形象地可比喻为"熟鸡蛋"; 而一般中子星只有表面的壳层才具有刚性(就像"生鸡蛋"). 刚体不论是自由转动或是在力矩作用下转动, 都容易产生进动. 包括 PSR B1821-11 以及其他一些脉冲星类在内的计时(Timing)和轮廓等观测特征暗示某些致密星很可能是具有整体刚性的. 关于脉冲星自转周期突然增加(Glitch)现象, 在固态夸克集团星模型中, 不论是常规还是缓慢自转突变原则上都能很好地理解. 星震是固态致密星模型中是非常自然的, 它同时导致引力能与弹性能的释放. 星震时释放的能量可足够维持软γ射线重复暴与反常 X 射线脉冲星的爆发活动.

我们注意到, 夸克集团星物态之所以硬源于夸克集团的非相对论性以及之间相互作用的"排斥芯". 众所周知, 极端相对论性物质的物态方程偏软; 也正因如此人们往往认为传统夸克星具有软的物态、低的极限重量. 然而, 夸克集团内部价夸克的数目往往大于 3, 可为 6, 12, 18 等, 其质量较大(大于重子质量)而相对于动能不可被忽略. 并且, 类似于核子之间的相互作用, 夸克集团之间也可能存在强的短程排斥. 模型计算也确实表明, 夸克集团星有很大的参数空间容许其极限质量超过 2$M_\odot$, 能自然地理解脉冲星 PSR J1614-2230 和 PSR J0348+0432 的质量测量值.

现今, 由压缩重子物质组成致密星的研究因这两颗近两倍太阳质量脉冲星的发现而再次成为热点: 如何使得这类致密物态足够"硬"以获得超过两倍太阳质量的极限重量? 考虑超子自由度时, 因超子的产生阈密度较低而易于在致密星内部形成, 但却使得物态趋软(即所谓的"超子之谜"). 另一方面, 如果考虑夸克自由度, 因强相互作用的渐近自由也会导致高密度较软的物态. 而我们提出的夸克集团星区别于常规的中子星和夸克星, 因其基本单元为非相对论运动的夸克集团而具有较硬的物态; 这一特征在两倍太阳质量脉冲星发现之前就已经预言. 诚然夸克集团存在的物理基础还需要进一步研究、论证, 但如前所述它并非一般意义上的非物理(Unphysical)概念. 尽管一些非物理模型可在大质量脉冲星发现之后建立起来以解释观测现象, 然而固态夸克集团星模型[25]在十年之后之所以还不被忽略就在于得到了若干后续观测的支持. 除了大质量脉冲星与硬的物态的发现, 还包括星震释能机制免除对 magnetar 模型的质疑、两类星震过程跟两类 glitch 观测的吻合等. 这些十年前未被意识到的特征及其未来更多观测检验, 体现了固态夸克集团星的生命力.

脉冲星表面普遍具有~$10^{12}$G 的磁场(相应磁矩$\mu$~5×$10^{29}$ G cm$^3$, 而电子自旋磁矩$\mu_e$=9×$10^{-21}$ G cm$^3$), 其起源一直没有定论. 因 s 夸克稍重于 u/d 夸克, 奇异物质中电子数密度约为夸克数密度~$10^{-5}$, 则一颗奇异夸克集团星内电子的总数 $N_e$~$10^{-5}$×$10^{57}$~$10^{52}$. 这些电子分布于~10 km 的球内, Fermi 动量 $P_F$~10 MeV/$c$. 然而, 将这些电子简单地看作理想费米子是过于粗

糙了. 很容易估计电子之间的 Coulomb 作用能 $E_c \sim 10^{-2}$ MeV. 因自旋空间交换对称性分布将降低电子之间的 Coulomb 能, Fermi 面附近约 $E_c/c$ 动量范围的电子可能倾向于具有一致的自旋磁矩. 这些电子的总数 $(4\pi p_F^2 E_c / c)/(4\pi p_F^3 /3)N_e \sim 10^{49}$, 它们贡献的磁矩总量 $\sim 10^{49} \mu_e \simeq 10^{29}$ G cm$^3$ 这与普通脉冲星磁矩的观测值相当. 所以, 固态夸克星的铁磁相变很可能是脉冲星磁场的一种起源. 当然相对论性电子气的铁磁相变机制还需要深入地探索. 我们期待未来这方面的理论研究进展.

此外, 夸克集团物态的唯像研究、脉冲星的星周环境及其观测影响、辐射行为等也值得关注. 例如, 可以借鉴类惰性原子的相图, 依据"态对应原理"(Corresponding-state Approach)探讨夸克集团之间的相互作用及其物态[31]; 考虑脉冲星周围遗迹盘的制动[32]所致自转行为的影响及演化; 热和非热 X 射线的偏振特征[33], 等等.

存在质量远小于"H 星"的"H 团块"(也是由 H 集团为基本自由度而构成的凝聚体, 只是质量要比 H 星甚至 H 行星低得多)吗? 对于夸克集团相的计算表明, 由夸克集团构成的小团块是可能存在的. 如果这些集团是 H 集团的话, 我们可以称它为 H 团块. H 团块或许存在于宇宙线中; 如何从观测到的宇宙线特殊事例中寻找 H 团块的证据是值得思考的.

## 4 脉冲星研究的未来

随着我国经济、社会等领域的发展, 反映国家关键实力的科学技术研究必将越来越受到重视; 脉冲星研究当然也不例外. 事实上, 我国正在和未来要建设的若干地面和空间望远镜项目正昭示着脉冲星研究的美好未来. 这些研究无疑将很大程度上提升国家科技实力、增强国际学术话语权, 而且也有利于检验前述的固态夸克集团星模型.

目前我国已经启动两个与脉冲星研究紧密相关的大科学工程, 即"500 m 口径球面射电望远镜"(简称 FAST)和"空间硬 X 射线调制望远镜"(简称 HXMT). FAST 利用我国贵州省黔南州平塘县大窝凼洼地的独特地形条件而建, 其接收面积相当于约 30 个足球场大小. 2016 年建成后, FAST 将是全球口径最大的单口径射电望远镜. 作为单天线, FAST 在脉冲星搜寻及监测方面的效率要优于天线阵, 其高的探测灵敏度有望发现上千颗射电辐射非常微弱的脉冲星. 利用 FAST 新发现的某些特殊类型的脉冲星将会加深人们对引力和引力波、致密物质状态、星际磁场和介质等方面的认识. 此外, FAST 刚建成后的低频脉冲星巡天对于望远镜调试和积累运行经验也是非常有帮助的.

HXMT 预计 2014 年底或 2015 年初发射, 它将是我国第一颗天文科学卫星. 它包括高能探测器(有效探测面积约 5100 cm$^2$, 覆盖能区: 20–250 keV), 中能探测器(952 cm$^2$, 5–30 keV)和低能探测器(384 cm$^2$, 1–15 keV), 对孤立或吸积脉冲星的非热 X 射线辐射敏感.

除了我国的 FAST 和 HXMT 外, 国际上与脉冲星科学紧密相关的几个大型设备也正在计划或建设中. (1) 平方公里阵(Square Kilometre Array, SKA)将是世界上最大、最灵敏的射电望远镜; 中国已经正式加入这一国际合作. 它是由 3000 个口径 15 m 的射电天线组成的五臂阵列干涉仪, 从阵列核心到边缘的距离达 3000 km. SKA 核心的候选台址位于南非和澳大利亚境内. (2) 低频射电阵(Low Frequency Array; LOFAR)着眼于打开宇宙的甚低频射电窗口(频段在 10–240 MHz), 主要建于荷兰的东北部, 也是由小的射电天线组成的干涉仪. 它将在技术和科学上支持 SKA. 2012 年 LOFAR 首次巡天. (3) 核谱望远镜阵(Nnuclear Spectroscopic Telescope Array, NuSTAR)能在硬 X 射线(频段 6–79 keV) 波段成像, 并已于 2012 年 6 月 13 日发射升空. (4) 中子星内部组分探险者(Neutron Star Interior Composition Explorer, NICER)为美国 NASA 计划建设的大面积聚焦的 X 射线计时望远镜, 可望在 2016 年放置于国际空间站, 能够得到 0.2–12 keV 能区的转动相位分离谱并进行 X 射线导航试验. (5) 除了电磁波外, 包括 LIGO 和 Virgo 在内的~kHz 引力波望远镜也有望给出单颗脉冲星持续引力波辐射信号, 甚至探测到超新星爆发和脉冲星双星合并时产生的引力波爆发事件. 这些引力波信息将进一步限制脉冲星物态.

值得一提的是, 除了基础科学的学术意义, 脉冲星研究还具有珍贵的应用价值; 后者主要体现在时间标准和航天器导航两方面. 尽管个别脉冲星的自转周期的长期稳定性已经赶上甚至超过了原子钟, 但利用单颗脉冲星计时还是不现实的. 未来要真正实现脉冲星时间标准, 需要建立脉冲星计时阵(Pulsar

Timing Array, PTA). 同样地, 脉冲星导航也离不开 PTA. 而利用 PTA 准确测量脉冲到达时间也将最终成就基础研究. 这至少体现在以下两个方面.

(1) PTA 探测宇宙低频背景引力波. 我们知道毫秒脉冲星具有比较稳定的自转周期, 但也存在到达时间的残差. 不同方位脉冲星到达时间残差的关联性起因迥异: 单极关联由时钟不精准引起, 偶极关联反映确定太阳系质心的偏差(其修正可改进木星质量的测量值), 而四极关联则是存在于脉冲星和地球之间的($10^{-9}$–$10^{-7}$ Hz)背景引力波所导致. 目前国际上几个活跃的脉冲星研究小组都把利用 PTA 测量背景引力波当作核心课题.

(2) 通过脉冲星计时行为研究脉冲星物态与结构, 并检验引力理论. 导致计时不稳定(如 Glitch, 噪声等)的因素很多, 部分因素与脉冲星物态与结构有关. 与致密物态相关的脉冲星内部结构的变化将导致自转周期随时间的内禀变化. 因此, 精确地监测脉冲星自转周期, 从复杂多变的计时行为中分辨出与物态相关的特征信号, 无疑有助于解决致密物态这一挑战性问题. 而对于双星系统中的脉冲星而言, 不同引力理论预言其不同的演化行为; 这些也可以通过脉冲星计时测量检验.

可见, 脉冲星的基础和应用研究紧密关联. 未来依托 FAST 等望远镜建立中国 PTA 并高精度计时若干脉冲星是我们期待的. 除了时间标准和导航方面的工程学意义, 脉冲星研究有助于了解致密物态、探测低频引力波. 当然, 个别脉冲星(特别是双星系统中的脉冲星)的高精度计时还能够检验引力理论基础(如引力常数的时变, 等效原理, Lorentz 对称性破缺性等). 总之, 借助越来越多高性能望远镜的运行, 脉冲星定将在展现其认识自然角色的同时服务于人类开拓自然的实践之中.

## 5 总结

80 年多前, Landau 为了解决恒星能量来源的问题, 综合了对引力束缚恒星的相关研究与 Gamow 原子核不可压缩液滴的猜想, 提出了巨大原子核的概念. 尽管因时代的局限性 Landau 的观点充斥着不少经不起历史推敲的错误概念, 但至今我们还不得不提及它: 因为"密度跟原子核相当的宏观物质"(即所谓的压缩重子物质)概念在那里被认真地考虑, 并且为了理解当今探测到的大量天文观测现象, 我们离不开这种物质. Landau 的概念逐步得到发展, 特别是在脉冲星

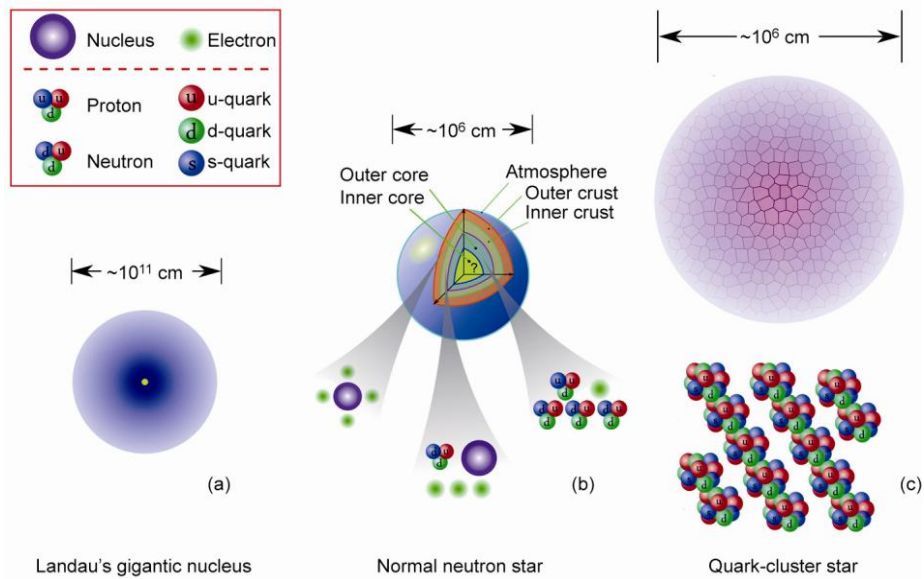

**图 3** 80多年前 Landau 推测的"中子"星(a)逐渐发展成当今流行的通常中子星(b); 而夸克集团星有别于前两者, 是夸克集团的凝聚体(c)

**Figure 3** Neutron star was speculated more than 80 years ago, which is then developed to normal neutron star of the mainstream. Quark-cluster star, as condensed matter of quark clusters, distinguishes from both neutron and conventional quark stars, which we expect to test further by future observations with advanced facilities.

发现之后, 成为目前较流行的中子星模型. 脉冲星是超新星爆发形成的压缩重子物质的主要形式. 而我们从天文物理角度出发, 为了理解各类脉冲星类致密天体的众多观测表现, 也提出在低温情形、几倍核物质密度下会形成夸克集团态. 这样, 脉冲星也因而可能是固态夸克集团星. 这些概念展示在图 3 中.

相对于中子星而言(图 3), 夸克集团星物态硬、整体刚性高、表面强自束缚, 这些都可能被超新星、伽玛暴、各类脉冲星等甚至宇宙线的观测检验. 我们高兴地看到, 国内或国际上若干重大天文观测设备正在启动或建设之中, 而脉冲星研究都是它们的关键科学之一. 至今观测未能原则上排除脉冲星类天体为夸克集团星, 其与流行中子星模型之间的竞争有望在未来一二十年内被观测裁定. 值得一提的是, 脉冲星研究的应用性也将有助于其科学目标的实现.

最后, 让我们借 Anderson P W(1923-)的一段名言来总结本文: "The ability to reduce everything to simple fundamental laws does not imply the ability to start from those laws and reconstruct the Universe." 对于几倍核物质密度下低温压缩重子物质状态的研究而言, 我们尤为感受尴尬和困惑: 那里的基本强相互作用性质(与七大千禧问题之一密切相关)依然还未确定, 更不用说"多体问题"了.

# Compressed baryonic matter: from nuclei to pulsars

XU RenXin[*]

*School of Physics and State Key Laboratory of Nuclear Physics and Technology, Peking University, Beijing 100871, China*

Our world is wonderful because of the negligible baryonic part although unknown dark matter and dark energy dominate the Universe. Those nuclei in the daily life are forbidden to fuse by compression due to the Coulomb repulse, nevertheless, it is usually unexpected in extraterrestrial extreme-environments: the gravity in a core of massive evolved star is so strong that all the other forces (including the Coulomb one) could be neglected. Compressed baryonic matter is then produced after supernova, manifesting itself as pulsar-like stars observed. The study of this compressed baryonic matter can not only be meaningful in fundamental physics (e.g., the elementary color interaction at low-energy scale, testing gravity theories, detecting nano-Hertz background gravitational waves), but has also profound implications in engineering applications (including time standard and navigation), and additionally, is focused by Chinese advanced telescopes, either terrestrial or in space. Historically, in 1930s, L. Landau speculated that dense matter at supra-nuclear density in stellar cores could be considered as gigantic nuclei (the prototype of standard model of neutron star), however, we address that the residual compact object of supernova could be of condensed matter of quark clusters. The idea that pulsars are quark-cluster stars was not ruled out during the last decade, and we are expecting to test further by future powerful facilities.

**pulsar, nuclear matter, quark matter, nucleus**